\documentclass[aps,prl,twocolumn,superscriptaddress,reprint,showpacs]{revtex4-2}
\usepackage{graphicx}
\usepackage{setspace}
\usepackage{amsmath,amssymb,amsfonts}
\usepackage{color}
\usepackage{array}
\usepackage{subfigure}
\usepackage{hyperref}
\usepackage{float}
\usepackage{lipsum}

\usepackage{braket}

\usepackage[all]{xy}
\newcommand{\RN}[1]{%
  \textup{\uppercase\expandafter{\romannumeral#1}}%
}

\begin{document}
\title{Sensing Low-Frequency Field with Rydberg Atoms via Quantum Weak Measurement}

\author{Ding Wang}
\thanks{Ding Wang, Shenchao Jin and Xiayang Fan contributed equally to this work.}
\affiliation{State Key Laboratory of Photonics and Communications, Institute for Quantum Sensing and Information Processing, School of Sensing Science and Engineering, Shanghai Jiao Tong University, Shanghai 200240, China}

\author{Shenchao Jin}
\thanks{Ding Wang, Shenchao Jin and Xiayang Fan contributed equally to this work.}
\affiliation{Shanghai Institute of Optics and Fine Mechanics, Chinese Academy of Sciences, Shanghai 201800, China}

\author{Xiayang Fan}
\thanks{Ding Wang, Shenchao Jin and Xiayang Fan contributed equally to this work.}
\affiliation{Shanghai Institute of Optics and Fine Mechanics, Chinese Academy of Sciences, Shanghai 201800, China}
\affiliation{University of Chinese Academy of Sciences, Beijing 100049, China}

\author{Hongjing Li}
\email[email: ]{lhjnet2012@sjtu.edu.cn}
\affiliation{State Key Laboratory of Photonics and Communications, Institute for Quantum Sensing and Information Processing, School of Sensing Science and Engineering, Shanghai Jiao Tong University, Shanghai 200240, China}
\affiliation{Hefei National Laboratory, Hefei 230088, China}
\affiliation{Shanghai Research Center for Quantum Sciences, Shanghai 201315, China}

\author{Jiatian Liu}
\affiliation{Shanghai Institute of Optics and Fine Mechanics, Chinese Academy of Sciences, Shanghai 201800, China}
\affiliation{University of Chinese Academy of Sciences, Beijing 100049, China}

\author{Jingzheng Huang}
\email[email: ]{jzhuang1983@sjtu.edu.cn}
\affiliation{State Key Laboratory of Photonics and Communications, Institute for Quantum Sensing and Information Processing, School of Sensing Science and Engineering, Shanghai Jiao Tong University, Shanghai 200240, China}
\affiliation{Hefei National Laboratory, Hefei 230088, China}
\affiliation{Shanghai Research Center for Quantum Sciences, Shanghai 201315, China}

\author{Guihua Zeng}
\affiliation{State Key Laboratory of Photonics and Communications, Institute for Quantum Sensing and Information Processing, School of Sensing Science and Engineering, Shanghai Jiao Tong University, Shanghai 200240, China}
\affiliation{Hefei National Laboratory, Hefei 230088, China}
\affiliation{Shanghai Research Center for Quantum Sciences, Shanghai 201315, China}

\author{Yuan Sun}
\email[email: ]{yuansun@siom.ac.cn}
\affiliation{Shanghai Institute of Optics and Fine Mechanics, Chinese Academy of Sciences, Shanghai 201800, China}
\affiliation{University of Chinese Academy of Sciences, Beijing 100049, China}

\begin{abstract}
Recently, Rydberg atom has emerged as an attractive choice to realize quantum sensing of low-frequency electric field. The progress so far has mostly utilized the intensity and phase changes in probe laser and the corresponding detection mechanism still remains classical. Nevertheless, external field acting on the Rydberg state can induce the polarization variation of probe laser in the Rydberg electromagnetically induced transparency (EIT) system embedded in realistic multi-state atoms. We experimentally observe this phenomenon and realize signal extraction by appropriately utilizing the polarization degrees of freedom. Based on such a mechanism, we further design and implement a quantum weak measurement scheme, which clearly suppresses the technical noise and leads to considerable improvement of performance. Evaluation of the sensitivities across different post-selection angles demonstrates that the weak measurement results agree well with the theoretical model predictions. The advantages of our method are analyzed from multiple aspects, including characterizing the responses over different frequencies and comparing the responses of the weak measurement scheme and the traditional transmission-based method. After accounting for the screening effect of a measured ratio 17\% where the $^\text{87}$Rb atoms experience a substantially reduced field inside the glass cell, the performance reaches 33 $\mu\text{V}~\text{cm}^\text{-1}~\text{Hz}^\text{-1/2}$ in sensitivity and 1.0 $\mu\text{V/cm}$ in minimal detectable field for an integration time of 1000 s, as perceived by the atoms.
\end{abstract}
\maketitle

Atomic ensembles are becoming increasingly important for detecting weak fields in the era of quantum sensing \cite{kominis_subfemtotesla_2003, budker_optical_2007, RevModPhys.82.1041, kitching_atomic_2011, degen_quantum_2017, lu_recent_2023}.
Among various applications, utilizing Rydberg atoms for electric field sensing has made solid progress and demonstrated exceptional potentials in radio-frequency (RF), microwave and THz for many years \cite{zhang_rydberg_2024}.
Such a remarkable capability stems from the giant polarizability of Rydberg atoms, which scales as $n^7$ with principal quantum number $n$ and confers high sensitivity to external electric fields \cite{adams_rydberg_2019, yuan_quantum_2023} that couple with transitions between atomic Rydberg states.
On the other hand, low-frequency (LF) electric fields, characterized by wavelengths exceeding $1~\text{km}$, are critically important in applications ranging from space science \cite{berthelier_ice_2006} and geology \cite{wang_assessment_2022} to communication in complex environments \cite{latypov_compact_2022}. 
Compared to traditional metal antennas, Rydberg atoms offer a compelling advantage that they can detect LF fields with a device of just several centimeters while maintaining sensitivity theoretically, making this approach attractive for a wide range of applications.
The mechanism of sensing LF fields does not rely on the resonant Rydberg-Rydberg transitions, and its recent experimental realization has exhibited several key strengths and started to garner significant interest \cite{jau_vapor-cell-based_2020, jin_heterodyne_2025}.

Through the electromagnetically induced transparency (EIT) effect, the information of electric field interacting with Rydberg states can be sensitively and often coherently imprinted onto the intensity or phase change of a probe laser \cite{RevModPhys.77.633, RevModPhys.82.1041, zhang_rydberg_2024}.  
In previous studies of Rydberg atom receivers, the polarization degrees of freedom remain largely unexplored and the high-sensitivity results mostly come from intensity or phase change of probe lasers. 
For instance, with respect to microwave signals some recent investigations start to employ polarization-based measurement for amplitude \cite{gomes_rydberg_2024} while the minimal detectable field still lags behind the state-of-the-art results \cite{yuan_quantum_2023}. 
On the other hand, no experimental studies adopting polarization measurement have been conducted for LF signals to date.
Nevertheless, the birefringence induced by the atom-light interaction can possibly provide an interesting opportunity for high-sensitivity measurement and a theoretical mechanism of polarization-induced interference has been proposed \cite{sun_polarization-induced_2018}.
The routines for detecting LF fields by intensity and phase changes in the probe laser have already been well-established, and the signal extraction process is still classical \cite{jau_vapor-cell-based_2020, jin_heterodyne_2025} so far.
As the known limitations in these methods call for new solutions, the physics of polarization in Rydberg EIT atom-light interaction awaits further exploration, and detecting polarization signal with advanced signal extraction approaches can hopefully provide unprecedented capabilities for sensing LF field with Rydberg atoms.

\begin{figure}[t!]
\centering
\includegraphics[width=0.49\textwidth]{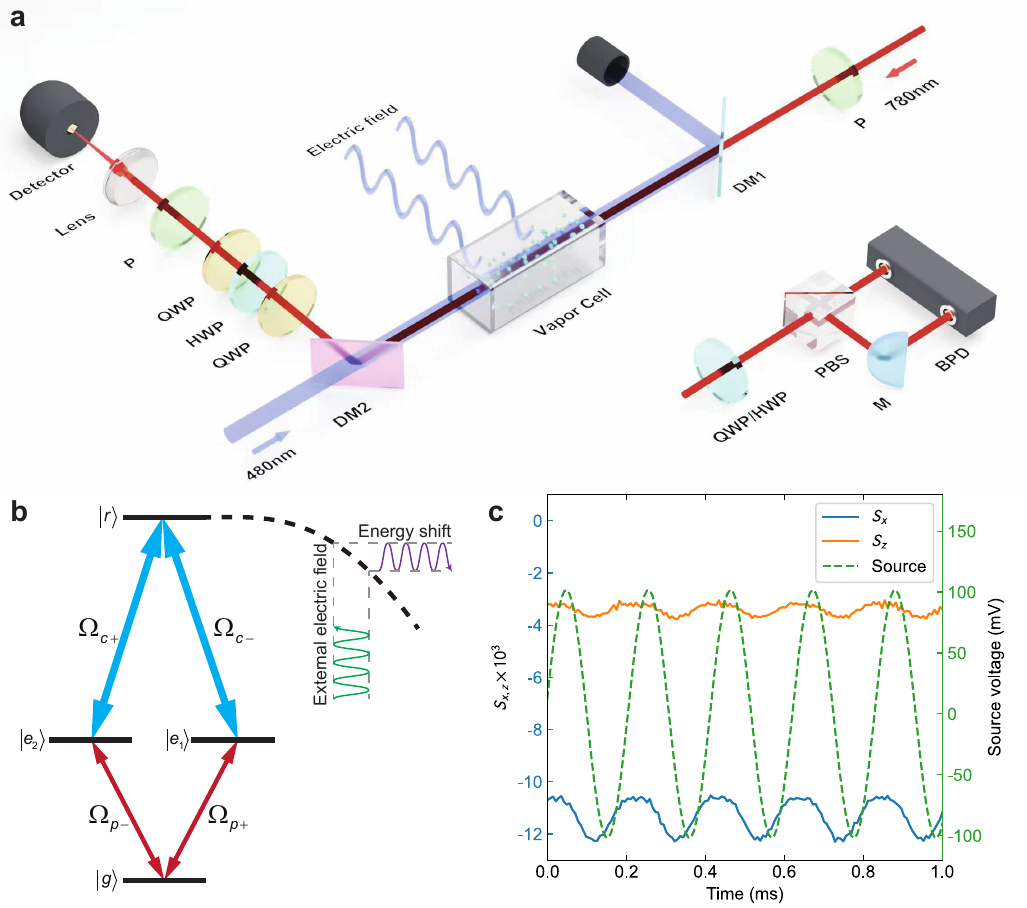}
\caption{\textbf{a,} Schematic diagram. \textbf{b,} Energy level. \textbf{c,} Polarization measurement result for the Stokes parameters $S_x$ and $S_z$. The abbreviations in the figure are as follows: P: polarizer, DM: dichroic mirror, HWP: half-wave plate, QWP: quarter-wave plate, PBS: polarization beam splitter, M: mirror, BPD: balanced photodetector.}
\label{fig1}
\end{figure}

In this work, we propose and realize a polarization-based Rydberg atom receiver for sensing kHz-frequency electric field, and we apply the method of weak measurement (WM) \cite{WA1, WA2} to enhance the signal-to-noise ratio (SNR), which is specifically tailored to extract information embedded in the polarization variation of probe laser. 
WM approach is known as a powerful tool for precision measurement \cite{WA2, WA4, Wang_2025, WA6, WA7, h2018, WA9, WA10} and capable of suppressing the technical noise in practical systems \cite{WA12}, which has been recently used to amplify the dispersion signals caused by MHz-frequency electric field in Rydberg atom receiver \cite{WA-RA}. 
Empowered by the quantum WM with respect to the polarization-based measurement, we experimentally observe the enhancement of SNR of up to 40 dB compared to the traditional transmission-based measurement. In particular, we achieve the sensitivity of $193 \pm 8~\mathrm{\mu V~cm^{-1}~Hz^{-1/2}}$ for an external 4.8 kHz signal. 
Furthermore, we characterize the responses over different frequencies and demonstrate that the polarization-based WM improves the robustness and stability of system, where an integration time of $1000~\mathrm{s}$ leads to a minimal detectable field of $6.1 \pm 0.9~\mu\mathrm{V/cm}$. 
The screening effect \cite{jau_vapor-cell-based_2020, lim_kilohertz-range_2023} due to the unavoidable atomic layer on the inner cell surface is also evaluated.
After accounting for the screening ratio of 17\% at 4.8 kHz, the performance reaches $33~\mathrm{\mu V~cm^{-1}~Hz^{-1/2}}$ in sensitivity and 1.0 $\mu$V/cm in minimal detectable field for an integration time of 1000 s with respect to the actual LF fields interacting with the atoms.

WM-assisted Rydberg atom LF electric field sensor is illustrated in Fig.~\ref{fig1}a, and the sensing procedure can be divided into four steps, i.e., initial polarization preparation of probe light, sensing with Rydberg atom sensing unit, post-selection and detection. During polarization preparation, the probe light is prepared in the $\pi/4$ linear polarization state, i.e., $\ket{\psi_i}=1/ \sqrt{2} (\ket{H}+\ket{V})$
, where $\ket{H}$ and $\ket{V}$ are horizontal and vertical polarization states, respectively. The sensing process takes place in a Rydberg-atom vapor cell, where a four-level atomic model is introduced to analyze the basic properties. Compared with the conventional three-level ground–excited–Rydberg linkage structure, this model includes an additional excited state, as depicted in Fig.~\ref{fig1}b. The atoms are driven into the Rydberg state by counter-propagating probe and coupling lasers. When an external LF electric field $E(t)$ is applied, the energy of the Rydberg state oscillates accordingly, which ultimately results in the polarization variation of the probe laser through a polarization-induced interference mechanism  embedded in the multi-state Rydberg EIT system\cite{sun_polarization-induced_2018}. After propagating through the vapor cell, the evolved state can be expressed as  
\begin{equation}
	\ket{\Psi_j}=\cos\chi(t) e^{\frac{-i\varphi(t)}{2}} |H\rangle + \sin\chi(t) e^{\frac{i\varphi(t)}{2}}|V\rangle,  
	\label{phi123}
\end{equation}
where $\varphi(t)=\beta E(t)$ represents the relative phase induced by  signal, with $\beta$ denoting the effective field-to-phase conversion coefficient, the linear relationship between $\varphi(t)$ and $E(t)$ holds strictly near two-photon resonance $\delta=0$, $\delta$ is the energy shift caused by $E(t)$, and $\chi (t)$ denotes higher-order components into the response. 

\begin{widetext}
As detailed in Supplementary S1, 
\begin{equation}
	\label{ana_tot}
	\begin{aligned}
		\beta &= \frac{\mathcal{A}k}{\Gamma_p}
		\frac{2(\Omega_{c-}^2-\Omega_{c+}^2) e^\mathcal{A}}
		{2\Omega_c^2 - (\Omega_{c-}+\Omega_{c+})^2 (1-e^{2\mathcal{A}})}, \\
		\chi(t) &= \frac{1}{2}\arccos(
		\frac{\Omega_{c-}^2-\Omega_{c+}^2}{\Omega_c^2}
		\frac{
			2\Omega_c^2 - (\Omega_{c-}+\Omega_{c+})^2(1+e^{2\mathcal{A}})
			+4\Omega_{c-}\Omega_{c+}e^{\mathcal{A}}
			\left[1-\frac{1}{2}\left(\frac{\mathcal{A}}{\Gamma_p}\right)^2 k^2 E^2(t)\right]
		}{2\Omega_c^2 - (\Omega_{c-}+\Omega_{c+})^2(1-e^{2\mathcal{A}})}),
	\end{aligned}
\end{equation}
where $\mathcal{A}$, $k$, and $\Gamma_p$ denote the total absorption, the first-order Stark polarizability, and power-broadened linewidth, respectively, $\Omega_{c\pm}$ are Rabi frequencies corresponding to $+, -$ components of the coupling laser, and $\Omega_c^2 = \Omega_{c-}^2 + \Omega_{c+}^2$ represents the total Rabi frequency.
\end{widetext}

During the post-selection process, the evolved probe laser is projected onto a predefined polarization state 
\begin{equation}
	\ket{\psi_f} =\frac{1}{\sqrt{2}}(ie^{ -i\varepsilon}\ket{H} - ie^{i\varepsilon}\ket{V}),
\end{equation}
which is nearly orthogonal to the initial polarization state with an extremely small post-selection angle of $\left| \varepsilon  \right| \ll 1$. To enable the real-time detection of LF electric field, the detected light intensity is acquired to extract the signal, which can be expressed as
\begin{equation}
	I(t)=I_0|\bra{\psi_f}\Psi_j\rangle|^2=\frac{I_0}{2}[1-\sin(2\chi(t))\cos(2\varepsilon- \beta E(t))],\label{I1}
\end{equation}
and $I_0$ is the initial light intensity. As derived in Supplementary S2, the application of post-selection will lead to a modified weak value of $A_w=\frac{\sin \varepsilon(\cos\chi(t)-\sin\chi(t))-i\cos \varepsilon(\cos\chi(t)+\sin\chi(t))}{\sin \varepsilon(\cos\chi(t)+\sin\chi(t))-i\cos \varepsilon(\cos\chi(t)-\sin\chi(t))} $, and Eq. \eqref{I1} can be further expressed as 
\begin{equation}
	I(t)\approx I_0 \xi [1+\beta E(t)\mathrm{Im}A_w], \label{I(t)}
\end{equation}
with $\xi =[1-\sin (2\chi(t)) \cos(2\varepsilon)]/2$, under the condition of $|\beta E(t) \mathrm{Im}A_w/2| \ll 1$ \cite{WA2}.

In practice,  the sensing is inevitably affected by various types of noise. 
The total noise amplitude, denoted as $\Delta I_{\mathrm{noise}}$, arises from the superposition of statistically independent noise sources. 
Here, we consider the noise amplitude with an integration time of 1~\text{s} and focus on three dominant terms in the system. 
The first is technical noise $\Delta I_{\mathrm{te}}$, such as relative intensity noise, device imperfection, and thermal noise \cite{WA4,WA11},
whose amplitude scales linearly with the optical intensity. 
The second is shot noise $\Delta I_{\mathrm{sh}}$ including photon shot noise and atomic projection noise. In a thermal atomic ensemble, the probe fluctuation induced by atomic shot-noise is negligible compared with the photon shot noise \cite{Atomicnoise}. We therefore retain only the photon shot noise contribution, and the amplitude is proportional to the square root of the light intensity.
The third is electronic noise $\Delta I_{\mathrm{el}}$, a constant noise floor contributed by the photodetector and detection circuits \cite{electric}, which is independent of the light intensity. 
Consequently, the total noise level is $\Delta I_{\mathrm{noise}} = \sqrt{(\Delta I_\mathrm{te}\xi)^2+(\Delta I_\mathrm{sh}\sqrt{\xi})^2+(\Delta I_\mathrm{el})^2}$ in our system.
The electric-field-induced signal amplitude at the first harmonic frequency is given by $I_{\mathrm{sig}} \approx I_{0}\xi\beta E(t)\mathrm{Im}A_w$ derived from Eq. \eqref{I(t)}, and SNR can be written as $\mathrm{SNR}=I_{\mathrm{sig}}/\Delta I_{\mathrm{noise}}$.
This indicates that the role of weak measurement becomes significant when the noise budget is dominated by technical noise, corresponding to a relatively high probe intensity. For nearly orthogonal pre- and post-selection states, i.e.,  $\varepsilon\ll 1$, the light intensity is reduced by a scaling factor $\xi \approx \sin^2 \varepsilon$, thereby mitigating technical noise, while field-induced response is less strongly suppressed due to the weak value $|\mathrm{Im}A_w |\approx \cot \varepsilon$, indicating that one can obtain an amplification factor of $|\mathrm{Im}A_w|$ on the extraction of $E(t)$. 
As probe light intensity decreases, the contributions of other noise become dominant, and the post-selection angle at the optimal SNR correspondingly shifts to larger values.

Prior to demonstrating the superiority of proposed scheme in LF electric field sensing, we characterize the response to LF field of the polarization-based measurement.
For this experiment, optical devices and photodiode located downstream of DM2 on the 780-nm optical path were changed to a balanced detection system,  comprising a half-wave or quarter-wave plate followed by a polarization beam splitter and a balanced photodetector, as shown in the lower right of Fig.~\ref{fig1}a.
Such a configuration follows the standard procedure of measuring the Stokes parameters $S_x$ and $S_z$ \cite{budker_sensitive_2000, zhang_dichroism_2021}.
A representative polarization signal is displayed in Fig.~\ref{fig1}c obtained under an applied electric field of $0.18~\mathrm{V/cm}$ amplitude at $4.8~\mathrm{kHz}$.
The results show that $S_x, S_z$ components exhibit a linear response to the applied field, with a slight phase delay attributable to the detection circuits.
This observation confirms the system's capability for LF electric field sensing using polarization-based, and then the WM-based detection technique.

The experimental setup is illustrated in Fig.~\ref{fig1}(a).
The core of the setup is an uncoated, enriched $^{87}$Rb vapor cell maintained at room temperature with the size of $25~\mathrm{mm}\times 25~\mathrm{mm}\times 50~\mathrm{mm}$.
A pair of $300~\mathrm{mm}\times 300~\mathrm{mm}$ copper plates, spaced $56~\mathrm{mm}$ apart, are positioned around the cell to generate the target AC electric field.
The influence of the geomagnetic field is suppressed by a set of 3-axis Helmholtz coils.
A ladder-type EIT scheme employing counter-propagating $780~\mathrm{nm}$ and $480~\mathrm{nm}$ laser beams is used to excite the Rydberg state.
The $780~\mathrm{nm}$ probe laser, with a beam waist of $0.3~\mathrm{mm}$, couples the ground state $|g\rangle = |5^{2}S_{1/2}, F=2\rangle$ to the excited states $|e_{1,2}\rangle$ of $|5^{2}P_{3/2}, F=3\rangle$.
The $480~\mathrm{nm}$ coupling laser, with a beam waist of $0.4~\mathrm{mm}$, drives the transition from $|e_{1,2}\rangle$ to the Rydberg state $|r\rangle = |58^{2}D_{5/2}\rangle$.
In the experiment, the probe light is initially prepared in a $\pi/4$ linearly polarization during the pre-selection stage, after which it feeds into the vapor cell.
The output light carrying polarization information is reflected by the dichroic mirror and subsequently passes through a half-wave plate to offset for the half-wave loss.
Then it enters the post-selection stage implemented by a quarter-wave plate (optical axis at $\pi/4$) followed by a polarizer oriented at $-\pi/4+\varepsilon$, where $\varepsilon = 0.19~\text{rad}$ (except for the data in Fig.~\ref{fig3}), and is finally collected by a photodiode.
Finally, the light is collected by a photodiode and recorded by a data acquisition card for further data processing, such as fast Fourier transform (FFT) analysis.

\begin{figure}[h]
\centering
\includegraphics[width=0.36\textwidth]{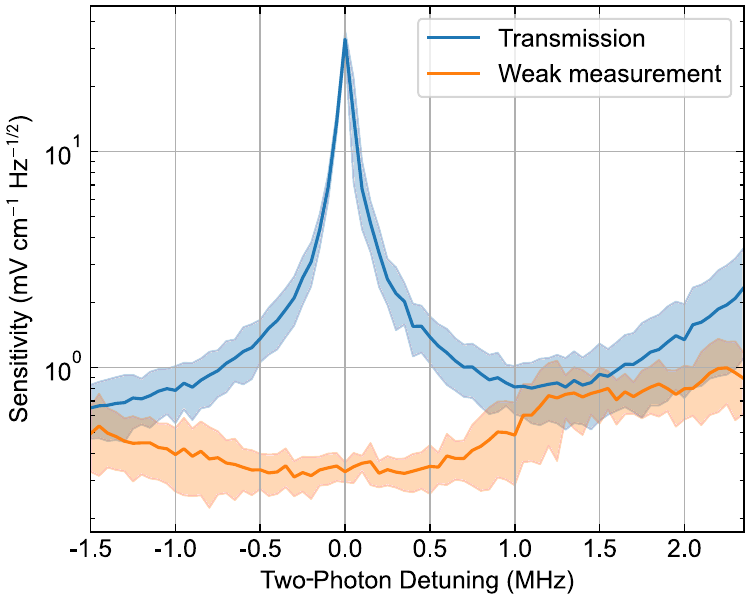}
\caption{Comparison between the traditional transmission measurement and the WM scheme. Scanning two-photon detuning of the Rydberg EIT is realized by scanning the frequency of 480 nm laser, which changes Rydberg atoms' response to LF field. The shaded regions represent one standard deviation derived from five independent measurements.}
\label{fig2}
\end{figure}

Previously, the prevailing approach for LF field sensing with Rydberg atoms is the transmission measurement method \cite{jau_vapor-cell-based_2020, li_super_2023, lei_high_2024}.
To benchmark the polarization-based WM scheme, we conduct a direct comparison with the traditional transmission method with the results summarized in Fig.~\ref{fig2}.
The traditional transmission measurement data are obtained by simply removing the polarizer preceding the photodiode.
To quantify sensitivity, we apply an electric field with an amplitude of $E = 0.18~\mathrm{V/cm}$ at $4.8~\mathrm{kHz}$.
The SNR is derived from the fast Fourier transform (FFT) spectrum of a $T=0.1~\mathrm{s}$ recorded trace, yielding a resolution bandwidth (RBW) of $1/T=10~\mathrm{Hz}$.
The sensitivity is then calculated as $E/ \mathrm{SNR}\cdot\sqrt{T}$.
According to Fig.~\ref{fig2}, the traditional transmission method suffers noticeable degradation in SNR near the two-photon resonance $\delta\approx 0$, which corresponds to the peak of Rydberg EIT spectrum.
In contrast, the WM-assisted Rydberg atoms system (see Methods) maintains high sensitivity around $\delta\approx 0$, manifesting the advantages of utilizing the polarization degrees of freedom ``with dispersion curve".
We observe the largest difference of SNR at $\delta=0$ that is about $40~\mathrm{dB}$.
It corroborates our analysis that the variation of the probe laser is sensitive to external field influencing the Rydberg EIT around $\delta=0$ as in Eq.~\eqref{ana_tot}, as the operation of polarization-based WM method is closely tied with the atom-laser interaction properties for sensing LF fields with Rydberg atoms. Moreover, while the sensitivity of traditional transmission method receives significant changes across exact resonance point due to the inherent Rydberg EIT properties,
the polarization-based WM scheme exhibits greater robustness against imperfections in controlling the value of two-photon detuning, which can become another advantage in practical applications.
 
\begin{figure}[h]
\centering
\includegraphics[width=0.36\textwidth]{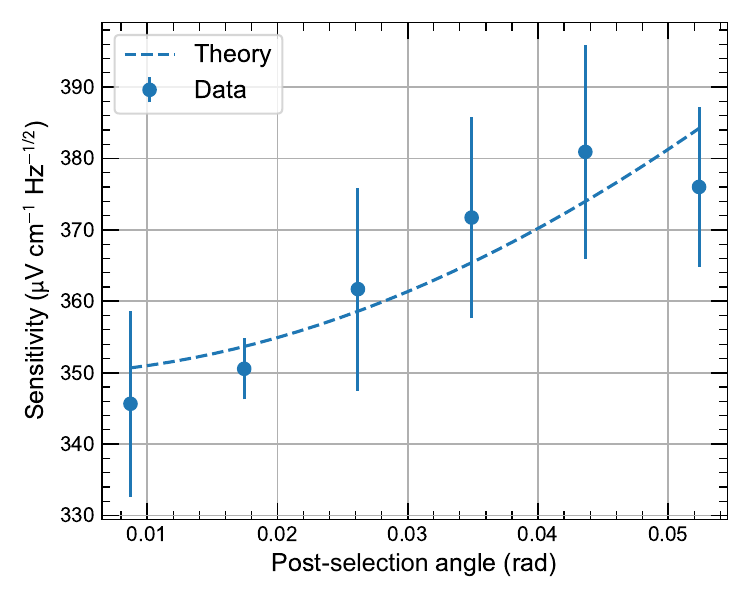}
\caption{Experimental and theoretical sensitivity as a function of the post-selection angle. The error bars represent one standard deviation from five independent measurements. The theoretical fitting is provided (see Supplementary S4) by modeling the post-selection process.
}
\label{fig3}
\end{figure}

We further investigate the sensitivity of the polarization-based WM scheme as a function of the post-selection angle to elucidate the basic properties, with the results shown in Fig.~\ref{fig3}.
For these measurements, the probe laser's Rabi frequency is set to $2\pi\times 50~\text{MHz}$.
Consistent with conventional WM experiments \cite{WA12, Wang_2025}, we observe that as the post-selection angle decreases, the scheme effectively suppresses classical noise, thereby improving the sensitivity.
In practice, non-ideal conditions such as imperfect EIT, velocity distributions within the hot atom ensemble, laser noise and the fluctuation of effective atom number participating the interaction introduce various noise contributions into the observables like $\varphi(t)$, $\chi(t)$ and $I_0$.
The overall noise can be categorized into three main types: classical noise such as relative intensity noise of light source, quantum noise including laser shot noise and atomic projection noise, and extra noise such as the photodiode electronics noise.
Owing to their distinct dependencies on laser power and post-selection angle, we employ a lower probe power in the following experiments together with an adjusted post-selection angle (see Methods) to optimize sensitivity.

\begin{figure}[h]
\centering
\includegraphics[width=0.49\textwidth]{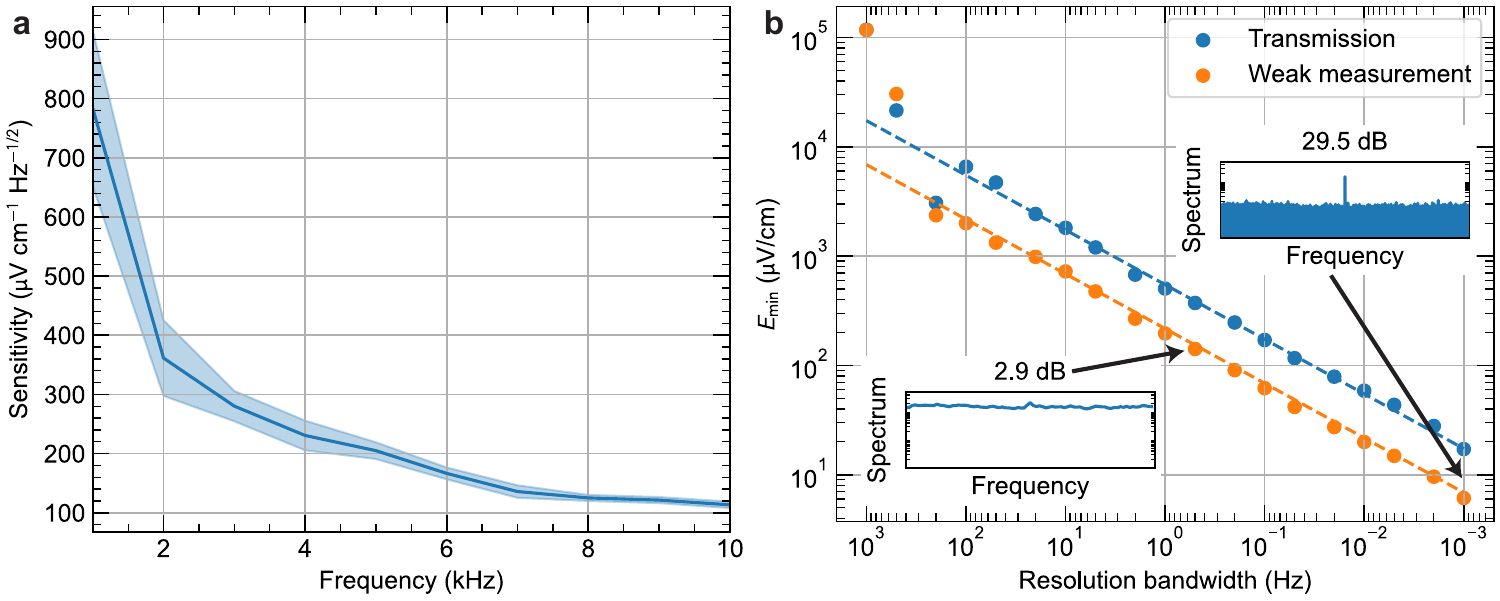}
\caption{\textbf{a,} Measured sensitivity across different AC electric field frequencies. The shaded region denotes one standard deviation from five independent measurements. \textbf{b,} Measured minimal detectable field strength versus RBW. The dashed lines are linear fittings according to the function $E_\mathrm{min} = k/\sqrt{T}$, where $k$ is the fitting parameter. Data are shown for a transmission measurement where the probe light at the photodiode was attenuated to match the optical power of the WM case. 
Insets: SNR extracted from the FFT spectrum for a $4.8~\mathrm{kHz}$, $0.18~\mathrm{mV/cm}$ electric field, with standardized y-axis limits for comparative purpose.}
\label{fig4}
\end{figure}

The sensitivity across different electric field frequencies up to $10~\mathrm{kHz}$ is characterized in Fig.~\ref{fig4}a.
The observed frequency dependence is attributed to the screening effect \cite{jau_vapor-cell-based_2020, lim_kilohertz-range_2023}, in which the free charges in the unavoidable atomic layer coated on the inner surface of cell respond to the external field.
This effect reduces the actual field strength reaching the atoms and becomes more severe for lower frequencies as the results suggest.
While the difference in frequency does not introduce fundamental difference in sensitivity as seen by the atoms, the system with currently available vapor cells is endowed with superior performance at relatively higher frequencies on practical accounts. 

The minimal detectable field is a major metric in quantum sensing of LF field. We carry out such an evaluation at an extended averaging time of $1000~\mathrm{s}$, with results shown in Fig.~\ref{fig4}b.
This field, calculated as $E_\mathrm{min} = E/ \mathrm{SNR}$, continues to decrease up to $T = 1000~\mathrm{s}$, reaching a value of $6.1 \pm 0.9~\mathrm{\mu V/cm}$ at $4.8~\mathrm{kHz}$ within the investigated parameter range, corresponding to a sensitivity of $193 \pm 8~\mathrm{\mu V~ cm^{-1}~ Hz^{-1/2}}$.
On a logarithmic scale, $E_\mathrm{min}$ exhibits a linear relationship with RBW $1/T$, consistent with $E_\mathrm{min}\propto 1/\sqrt{T}$, except at RBW values above $100~\mathrm{Hz}$ where the broadening bandwidth incorporates multiple technical noise peaks at lower frequencies.
For comparison, the minimal detectable field for the traditional transmission measurement is also plotted in Fig.~\ref{fig4}b, clearly demonstrating the advantage of WM.

It is important to note that the electric field values reported above correspond to the externally applied field.
The actual field experienced by the atoms is attenuated as a result of the screening effect.
By measuring this effect at $4.8~\mathrm{kHz}$ through the second-order Stark effect, we determine a screening ratio of $\eta = 17\%$ (see Supplementary S3).
This implies that the effective field sensed by the atoms is $E_\mathrm{eff} = \eta E$.
After accounting for this attenuation, the intrinsic sensitivity of the Rydberg atoms themselves is $33~\mathrm{\mu V~ cm^{-1} ~Hz^{-1/2}}$, with an effective minimal detectable field of $1.0~\mathrm{\mu V/cm}$.

The screening effect constitutes a major limitation for lower frequencies and future work plans to replace the conventional glass cell with a sapphire cell to alleviate this issue.
On the other hand, the achieved sensitivity remains far from the standard quantum limit (SQL) in our experiment, calling for further systematic investigations. 
For instance, the choice of the Rydberg state warrants careful evaluation for the optimal principal quantum number $n$ to enhance sensitivity.
Additionally, the currently deployed readout for linear response of $E(t)$ is valid only under the weak coupling condition $|\beta E(t) \mathrm{Im} A_w/2| \ll 1$,
while stochastic field fluctuations or large signal excursions may violate this condition in realistic scenarios, resulting in nonlinearities and systematic bias.
Incorporating a controllable reference phase with closed-loop feedback could maintain operation within the linear regime, enabling a tunable sensitivity-dynamic range trade-off and improved robustness for practical applications.


In conclusion, we have successfully demonstrated the implementation of a WM scheme for LF electric field sensing with Rydberg atoms.
Our results underscore the significant potential of leveraging polarization degrees of freedom within this framework.
By replacing the conventional transmission measurement with our WM approach, we achieved an SNR improvement of up to $40~\mathrm{dB}$.
This technique effectively suppresses classical noise, as evidenced by the attained minimal detectable field, thereby confirming its capability for probing weak electric fields.
While this work focuses on a standard WM configuration, the methodology is readily compatible with advanced techniques such as power recycling \cite{lyons_power-recycled_2015, krafczyk_enhanced_2021}, promising further enhancements in sensitivity for future low-frequency electrometry.
Moreover, this WM protocol can be directly adapted to the Rydberg microwave sensing field, creating a novel synthesis with existing polarization spectroscopy \cite{gomes_rydberg_2024} for a more versatile and powerful sensing paradigm.

\section*{Acknowledgments}
\label{sec:acknowledgments}

The authors gratefully acknowledge funding supports from the National Key R\&D Program of China (Grant No. 2024YFB4504002), the Science and Technology Commission of Shanghai Municipality (Grant No. 24DP2600202), the National Natural Science Foundation of China (Grant No. 92165107, No.62471289, No. 61671287, No.61631014, and No. 61901258), the fund of the State Key Laboratory of Advanced Optical Communication Systems and Networks, Natural Science Foundation of Shanghai (Grant No. 24ZR1432900), Innovation Program for Quantum Science and Technology (Grant No.2021ZD0300703), and Shanghai Municipal Science and Technology Major Project (Grant No. 2019SHZDZX01). Shenchao Jin acknowledges support from the China Postdoctoral Science Foundation (Grant No. 2024M753359).

\bibliographystyle{apsrev4-2}

\renewcommand{\baselinestretch}{1}
\normalsize

\bibliography{sn-bibliography}
\end{document}